\documentclass[preprint,amsmath,amssymb,aps,nofootinbib,tightenlines]{revtex4}

\usepackage{graphicx}
\usepackage{subfigure}

\newcommand{\wt}{\widetilde}
\newcommand{\ovl}{\overline}

\newcommand{\be}{\beta}

\newcommand{\la}{\lambda}

\newcommand{\beq}{\begin{equation}}
\newcommand{\eeq}{\end{equation}}

\newcommand{\bea}{\begin{eqnarray}}
\newcommand{\eea}{\end{eqnarray}}
\newcommand{\nl}{\nonumber\\}
\renewcommand{\l}{\left}
\renewcommand{\r}{\right}

\begin{document}

\preprint{\begin{tabular}{l}
\hbox to\hsize{hep-ph/0312101 \hfill KIAS-P03090}\\
\end{tabular} }

\title{CP violation in $B \to \phi K_S$  decay at large $\tan\be$}

\author{Seungwon Baek}
\affiliation{School of Physics, KIAS, 207-43 Cheongryangri-dong, 
    Seoul 130-722, Korea }

\date{\today}

\thispagestyle{empty}

\begin{abstract}
We consider the chargino contribution to the CP violation in $B \to \phi K_S$ 
decay in the minimal supersymmetric standard model at large $\tan\be$. 
It is shown that the Wilson coefficient $C_{8g}$ of the chromomagnetic 
penguin operator can be significantly enhanced by the 
chargino-mediated diagrams
while satisfying other direct/indirect experimental constraints.
The enhanced $C_{8g}$ allows large deviation in the CP asymmetry
from the standard model prediction.
\end{abstract}

\pacs{PACS numbers:12.60.Jv,11.30.Er,13.20.He}

\maketitle

%%%%%%%%%%%%%%%%%%%%%%%%%%%%%%%%%%%%%%%%%%%%%%%%%%%%%%%%%%%%%%%%%%%%%%%%%%%%%
The measurement of CP asymmetries at B-factories is a powerful
probe of new physics (NP). 
In the standard model (SM), the origin of CP violation is the
phase $\delta_{\rm CKM}$ in the $3\times 3$ Cabibbo-Kobayashi-Maskawa (CKM) 
matrix of quark mixing. In general, there can be many new sources of
CP violation in new physics models beyond the SM.

The time-dependent CP asymmetry in the neutral $B$ decays to CP eigenstates
$B \to f_{\rm CP}$
gives information on the two classes of CP violation $C_f$ and 
$S_f$~\cite{Nir:2002gu}:
\bea
  A_{\rm CP}(t) &=&\frac{\Gamma(\ovl{B}(t)\to f_{\rm CP})
                            -\Gamma({B}(t)\to f_{\rm CP})}
  {\Gamma(\ovl{B}(t)\to f_{\rm CP})+\Gamma({B}(t)\to f_{\rm CP})} \nl
  &=& -C_f \cos (\Delta m_B t) + S_f \sin (\Delta m_B t),
\label{eq:asym}
\eea
where $\Delta m_B$ is the mass difference of the $B$ system and
$B(\ovl{B})(t)$ is the state at time $t$ which
was $B(\ovl{B})$ at $t=0$. The CP asymmetries $C_f$ and $S_f$
are determined by
\bea
  \la_{\rm CP} \equiv e^{-2i(\be+\theta_d)} \frac{\overline{A}}{A},
\label{eq:lambda}
\eea
where $\be (\theta_d)$ is the contribution of SM (NP) to 
the phase in the $B-\ovl{B}$ mixing
and $\ovl{A}(A)$ is the decay amplitude for  $\ovl{B}(B)\to f_{\rm CP}$.

While the time-dependent CP asymmetry in $B\to J/\psi K$ decay
\bea
 \sin 2\be_{J/\psi K_S} \equiv S_{J/\psi K_S} = 0.734 \pm 0.054.
\label{eq:sin2b_exp}
\eea
is consistent with the CKM picture of CP violation in the SM,
the CP asymmetry in $B\to \phi K_S$, $S_{\phi K_S}$, measured
at Belle~\cite{belle03} 
\bea
S_{\phi K_S} = -0.96 \pm 0.50 ^{+0.09}_{-0.11}
\label{eq:belle03}
\eea
differs from the SM expectation by 
$3.4 \sigma$\footnote{The result of BaBar~\cite{babar03} is consistent with
the SM at $1 \sigma$ level,
\bea
S_{\phi K_S} = 0.45 \pm 0.43 \pm 0.07.
\eea}.
With the presence of NP,
large deviation from $\overline{A}/A = 1$ is generally allowed
in the loop-induced process $B \to \phi K_S$
contrary to the tree-level diagram dominated $B\to J/\psi K_S$ decay.
Then from (\ref{eq:lambda}) we can see (\ref{eq:belle03}) can be
generated with  (\ref{eq:sin2b_exp}) untouched in case
$\overline{A}/A$ has a large phase.

The minimal supersymmetric standard model (MSSM) is one of
the most promising candidates of NP.
In the MSSM, it has been shown that (\ref{eq:belle03}) can be accomodated
if there exist new flavor structures in the
up- or down-type squark mass 
matrices~\cite{b2kp_glu}.
In this talk we show 
the phases in the flavor conserving soft parameters
($\mu, A_t, M_2$) 
may be responsible for the deviation in (\ref{eq:belle03}) via 
chargino-scalar top loops~\cite{Baek:2003kb}.
In this case the source of flavor mixing is CKM as in the SM.

Specifically, we adopt a decoupling scenario where the masses of
the first two generation scalar fermions are very heavy 
($\gtrsim {\cal O}(10 \mbox{ TeV})$),
so that the SUSY FCNC and SUSY CP problems are solved
without a naturalness problem.

In this decoupling scenario, it has been shown that 
the light stop and chargino contributions still allows
large direct CP asymmetry in the radiative
$B$ decay up to $\pm 16$\%~\cite{Baek:1998yn}.
It should be noted that the new contribution to $B-\ovl{B}$ mixing 
is very small and this
model is naturally consistent with 
(\ref{eq:sin2b_exp})~\cite{Baek:1998yn}.

In the analysis we have decoupled the charged Higgs contribution by assuming
$m_{H\pm} \gtrsim 1$ TeV to evade the very stringent two-loop EDM
constraints at large $\tan\be$ and relatively light pseudo 
scalar Higgs~\cite{Chang:1998uc}. 

We note that the $C_{7\gamma(8g)}^{\chi^\pm}$ has the enhancement
factors $m_{\wt{\chi}_I^\pm}/m_b$ by the chirality flip inside the loop
and can dominate the SM contribution.
On the other hand the Wilson coefficients of QCD penguin operators
,$C_{3,\cdots, 6}$, preserve the chirality 
and don't have such enhancement factors. 
In addition, due to the super-GIM mechanism,
the chargino contribution to $C_{3,\cdots, 6}$ are much smaller than the SM
values.
The contribution to chirality flipped operators 
$C_{7\gamma(8g)}^{'\wt{\chi}^\pm}$
are suppressed by $m_s/m_b$.
We neglect the contribution of electroweak penguin operators
which are also expected to be negligible in our scenario. 
Therefore the large
deviation in $\overline{A}/A$ should be generated soley by $C_{8g}$
in this scenario.

To calculte the hadronic matrix elements we use the QCD factorization
method in 
ref.~\cite{Beneke:2001ev}. 
The required $C^{\wt{\chi}^\pm}_{8g}(m_b)$ 
($C^{\rm SM}_{8g}(m_b)\approx -0.147$) 
for large deviation of CP asymmetries
can be estimated from the approximate
numerical expression for $\overline{A}$:
\bea
 \ovl{A} &\propto& 
  \sum_{p=u,c} V_{ps}^* V_{pb} (a_3 + a_4^p + a_5) \nl
  &\approx& -3.9 \times 10^{-4} ( 3.7 e^{0.21 i} + 4.5 C_{8g} ).
\label{eq:approx}
\eea
In the above expression the largest errors  which are about 20\% 
come from the decay constants
and form factors. These uncertainties are largely canceled 
in the asymmetries in (\ref{eq:asym}).

In (\ref{eq:approx}) and the following analysis, we did not include
weak annihilation contributions which are power suppressed but may be
numerically important. 
These contributions are not derived in QCD factorization on first principles,
and additional phenomenological parameter which can contain strong phase
should be introduced to estimate the divergent integral
at end point. 
The CP asymmetries $C_{\phi K}$, $S_{\phi K}$
can be increased depending on the size and phase of the parameter.

\begin{figure}
\begin{center}
\includegraphics[width=0.45\textwidth]{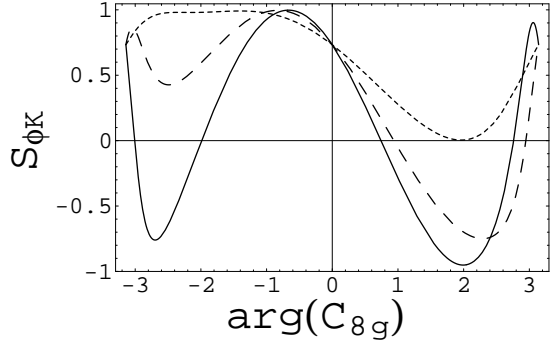}
\end{center}
\caption{$S_{\phi K}$ as a function of $\arg(C_{8g})$ for 
$|C_{8g}|=$0.33 (short dashed line), 0.65 (long dashed line) 
and 1.0 (solid line). }
\label{fig:Skp-C8}
\end{figure}
In Fig.~\ref{fig:Skp-C8}, we show $S_{\phi K}$ as a function of $\arg(C_{8g})$
for $|C_{8g}|=0.33,0.65$ and $1.0$.
From this figure, we can see $|C_{8g}| \approx 0.33-0.65$
with large positive phase can accomodate the deviation within 1$\sigma$.

Since the chargino contributions to $C_{7\gamma}$ and $C_{8g}$
have similar structures, one may think that large deviation in $C_{8g}$ 
may result in the large deviation $C_{7\gamma}$. 
Too large deviation in $C_{7\gamma}$
will violate the measurement of $B(B\to X_s \gamma)$,
for which we take 
$2\times 10^{-4} < B(B\to X_s \gamma) < 4.5 \times 10^{-4}$~\cite{Chen:2001fj},
because it is already consistent with the SM prediction.

Due to the different loop functions, however, the two Wilson coefficients
are not strongly correlated, and it is possible to have large deviation
in $C_{8g}$ while keeping $|C_{7\gamma}|$ constrained to satisfy 
$B(B\to X_s \gamma)$.
Since $C_{8g}$
is proportional to $\tan\be$ for large $\tan\be$, 
we need relatively large $\tan\be$
to have sizable effects.

The neutral Higgs boson $h^0$ which is lighter than the $Z$-boson at
tree level has large radiative corrections~\cite{Okada:1990vk} 
which can be approximated at large $\tan\be$ as
\bea
  m_{h^0}^2 \approx m_Z^2 + {3 \over 2 \pi^2} \frac{m_t^4}{v^2}
         \log \l(m_{\wt{t}_1} m_{\wt{t}_2} \over m_t^2\r),
\eea
where $v \approx 246$ GeV.
LEP II sets the lower limit on the SM-like Higgs boson mass:
$m_{h^0} \gtrsim 114.3$ (GeV) at 95\% CL~\cite{Hagiwara:fs}. 
This limit gives very strong constraint on the stop masses.

At large $\tan\be$($\sim {\cal O}(50)$), the SUSY QCD and SUSY electroweak
correction to the non-holomorphic couplings 
$H_u^* D^c Q$ become very important.
This gives large correction on the down-type quark masses
and some CKM matrix elements~\cite{Hall:1993gn}
\bea
  m_b = \frac{\ovl{m}_b}{1+\wt{\epsilon}_3 \tan\be},\quad
  V_{JI} = V_{JI}^{\rm eff} 
  \l[\frac{1+\wt{\epsilon}_3 \tan\be}{1+\epsilon_0 \tan\be}\r]
\eea
where $\wt{\epsilon}_3 \approx \epsilon_0 + \epsilon_Y y_t^2$
and $(JI)=(13)(23)(31)(32)$.
$\ovl{m}_b$, $V_{JI}^{\rm eff}$ are $b$-quark mass and
CKM elements measured at experiments, respectively.
Note that these SUSY threshold corrections are not
easily decoupled even for very heavy superparticles. %$m_{\wt{g}}$.
For degenerate SUSY parameters, $\epsilon_0 \approx 0.013$
and $\epsilon_Y \approx 0.003$.

\begin{figure}
\centering
%\subfigure[]{
%\includegraphics[width=0.43\textwidth]{tb35-50000.eps}
%}
%\hspace{-0.8cm}
%\subfigure[]{
%\includegraphics[width=0.43\textwidth]{tb60-50000.eps}
%}
\includegraphics[width=0.6\textwidth]{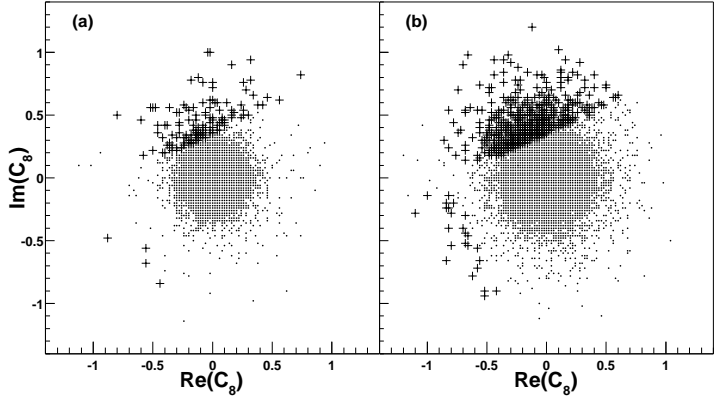}
\caption{Distribution of $C_8$ values for $\tan\be=35$ (a)
and $\tan\be=60$ (b) in the complex plane. 
See the text for other parameters.
The points with $S_{\phi K}<0$ are marked with $+$ symbol.}
\label{fig:C8}
\end{figure}

For the numerical analysis 
we fix the CKM matrix by using 
$|V^{\rm eff}_{us}|=0.2196$, $|V^{\rm eff}_{cb}|=4.12\times 10^{-2}$ and 
$|V^{\rm eff}_{ub}|=3.6\times10^{-3}$~\cite{Hagiwara:fs},
leaving only $\delta_{\rm CKM}$ as free parameter.
For example, for $\delta_{\rm CKM} = \pi/3$, we get 
$\Delta m_{B_d} = 0.491$ ps$^{-1}$ and $\sin2\beta=0.729$,
which are close to the experimental central values~\cite{Hagiwara:fs}.
%The $\delta_{\rm CKM}$
%is constrained by (\ref{eq:sin2b_exp}).
The free SUSY parameters in our scenario are
$\tan\be$, $M_2$, $\mu$, $m_{\wt{Q}}$, $m_{\wt{t}}$ and $A_t$
($m_{H^\pm}=1$ TeV)
(also $m_{\wt{g}}, m_{\wt{b}}$ are relevant for large $\tan\be$).
In Fig.~\ref{fig:C8}, we show the distribution of $C_{8g}$ in the complex
plane for $\delta_{\rm CKM}=\pi/3$, $\tan\be =35 (60)$, $m_{H^\pm}=1$ TeV,
$ m_{\wt{Q}} = 0.5 $ TeV, $m_{\wt{g}} =1$ TeV, and $m_{\wt{b}_R} = 0.5$ TeV.
We scanned the other parameters as follows: 
\bea
&& 0< m_{\wt{t}} < 1 \mbox{ TeV}, \quad 0 < |\mu| < 1 \mbox{ TeV}, \nl
&& 0< |A_t| < 2 \mbox{ TeV}, \quad 0 < |M_2| < 1 \mbox{ TeV}, \nl
&& -\pi < \arg(\mu),\arg(A_t),\arg(M_2) < \pi.
\eea
We have fixed the phase on the gluino mass parameter such that 
$\arg(m_{\wt{g}})+\arg(\mu)=\pi$
to maximize the SUSY QCD correction.
When scanning, we imposed the $B(B\to X_s \gamma)$ constraint
and the direct search bounds on the (s)particle masses~\cite{Hagiwara:fs}:
$m_{h^0} \ge 114.3$ GeV and
$m_{\wt{\chi}^\pm_1}, m_{\wt{t}_1}\gtrsim 100$ GeV.

From Fig.~\ref{fig:C8}, we can see that our scenario can easily
accommodate the discrepancy (\ref{eq:belle03}).
As mentioned above, we have chosen large value for the charged Higgs mass 
$m_{H^\pm}=1$ TeV to 
safely suppress the Barr-Zee type 
two-loop EDM constraints which are significant 
if $\tan\be$~\cite{Chang:1998uc} is large
and the pseudo-scalar Higgs boson ($A^0$) is relatively 
light~\footnote{The SUSY contribution
to $(g-2)_\mu$ is also small in this 
case.~\cite{Arhrib:2001xx}.}. We have checked that for the smaller $m_{H^\pm}$
the larger deviation in $S_{\phi K}$ is possible due to the cancellation
in the real part with the chargino contribution.
Therefore Fig.~\ref{fig:C8} is a rather conservative result 
for $S_{\phi K}$. We have checked that the allowed range for $S_{\phi K}$
is not sensitive to the change in $\delta_{\rm CKM}$ if we impose the
constraint (\ref{eq:sin2b_exp}).

The $B(B\to \phi K)$ varies moderately over the parameter space we considered
and seldom exceeds $15 \times 10^{-6}$, which is acceptable compared with
the experimental measurements. Also the mass
difference $\Delta m_s$ in the $B_s-\ovl{B_s}$ system is close to the
SM expectation $\Delta m_s \sim 14.5$ ps$^{-1}$ in the most region of parameter
space we considered, which may distinguish our scenario from other
scenarios in refs. \cite{b2kp_glu}.

For large $\tan\be$ in the MSSM, through the Higgs-mediated FCNC the 
$B(B_s \to \mu^+\mu^-)$ can be enhanced by a few orders of magnitude
over the SM prediction~\cite{bsll}. Observation of this leptonic decay mode,
for example at Tevatron Run II,
would be clear signal of NP.
However, this is possible only for relatively light $A^0$.
Since the large deviation in the $S_{\phi K}$,
although it needs new CP violating phase(s) of ${\cal O}(1)$,
does not necessarily require light $A^0$ as we have shown,
these two decay modes can be complementary
to each other in searching for the MSSM at large $\tan\be$.

In conclusion we have considered the chargino contribution to the CP 
asymmetries $S_{\phi K}$ and $C_{\phi K}$ of $B \to \phi K_S$ decay
in the CP violating MSSM scenario at large $\tan\be$.
We have shown that through the enhanced Wilson coefficient $C_{8g}$ 
of chromo-magnetic penguin operator by the large SUSY threshold 
corrections~\cite{Hall:1993gn}, there can be large deviations in the
CP asymmetries.

%\vfil\eject
\end{document}